# Electrically driven compact hybrid lithium niobate microring laser

JUNXIA ZHOU,[1,2] TING HUANG,[2] ZHIWEI FANG,[2,8] RONGBO WU,[2,9] YUAN ZHOU,[3,4] JIAN LIU,[2] HAISU ZHANG,[2] ZHENHUA WANG,[2] MIN WANG,[2] AND YA CHENG[1,2,3,5,6,7,*]

[1]State Key Laboratory of Precision Spectroscopy, East China Normal University, Shanghai 200062, China
[2]The Extreme Optoelectromechanics Laboratory (XXL), School of Physics and Electronic Science, East China Normal University, Shanghai 200241, China
[3]State Key Laboratory of High Field Laser Physics and CAS Center for Excellence in Ultra-intense Laser Science, Shanghai Institute of Optics and Fine Mechanics (SIOM), Chinese Academy of Sciences (CAS), Shanghai 201800, China
[4]Center of Materials Science and Optoelectronics Engineering, University of Chinese Academy of Sciences, Beijing 100049, China
[5]Collaborative Innovation Center of Extreme Optics, Shanxi University, Taiyuan 030006, China.
[6]Collaborative Innovation Center of Light Manipulations and Applications, Shandong Normal University, Jinan 250358, People's Republic of China
[7]Shanghai Research Center for Quantum Sciences, Shanghai 201315, China
[8] zwfang@phy.ecnu.edu.cn
[9] rbwu@siom.ac.cn
* ya.cheng@siom.ac.cn

**We demonstrate an electrically driven compact hybrid lithium niobate microring laser by butt coupling a commercial 980-nm pump laser diode chip with a high quality $Er^{3+}$-doped lithium niobate microring chip. Single mode lasing emission at 1531 nm wavelength from the $Er^{3+}$-doped lithium niobate microring can be observed with the integrated 980-nm laser pumping. The compact hybrid lithium niobate microring laser occupies the chip size of 3 mm × 4mm × 0.5 mm. The threshold pumping laser power is 6 mW and the threshold current is 0.5 A (operating voltage 1.64 V) in the atmospheric temperature. The spectrum featuring single mode lasing with small linewidth of 0.05 nm is observed. This work explores a robust hybrid lithium niobate microring laser source which has potential applications in coherent optical communication and precision metrology.**

Featured with its wide transparent window, high refractive index, large acusto-optic, electro-optic coefficients, and nonlinear optical coefficients, the integrated thin film lithium niobate (TFLN) photonics have emerged as a promising material platform for the fabrication of high-performance integrated photonic devices both for classical and quantum applications, such as low loss waveguide, high quality microresonator, high speed modulator and high efficiency optical frequency converters [1–6]. However, the single crystal lithium niobate lacks ability for efficient light generation as well as detection. Very recently, doping LN crystals with rare earth ions is an option to achieve the optical gain function in the TFLN platform, and optically pumped micro-lasers on TFLN have been demonstrated based on the rare earth ions doped TFLN [7–20]. Furthermore, electrically pumped hybrid lithium niobite/ semiconductor lasers have also been demonstrated by integration of the III-V laser on TFLN [21–25]. Several different complex integration approaches towards this goal have been developed, such as transfer printing, hybrid integration, flip-chip bonding and heterogeneous integration [21–25]. Such approaches all need advanced nanofabrication, micro-transfer and bonding techniques as well as rich experience in relevant areas. Compared with the III-V laser diode, the rare earth ions doped laser has wide bandwidth, polarization insensitivity, high temperature applicability, and good compatibility [26]. Furthermore, the rare earth ions doped laser is more promising to achieve high-power output, mode-locked operation and coherent beam combination [27, 28]. But now all of the rare earth ions doped TFLN lasers are optically pumped by external lasers using the fiber connection and hinder the development of integrated photonics on TFLN.

Here, we demonstrate for the first time an electrically driven compact hybrid lithium niobate microring laser by butt-coupling a 980-nm pump laser diode chip with a high quality $Er^{3+}$-doped lithium niobate microring chip rather than the fiber connection. The 980-nm pump laser diode is a commercial Chip-on-Submount (CoS) laser diodes, and the $Er^{3+}$-doped TFLN microring (Er: TFLN) laser is fabricated by the photolithography assisted chemo-mechanical etching (PLACE) technique in our lab [17]. Single mode C-band lasing emission form the Er: TFLN microring can be observed with the integrated 980-nm laser pumping. The compact hybrid lithium niobate microring laser occupies the chip size of 3 mm × 4 mm × 0.5 mm. The threshold pump laser power is 6 mW and the threshold current is 0.5 A (operating voltage 1.64 V) in the atmospheric temperature. The spectrum with single mode lasing and small linewidth of 0.05 nm is observed. This work explores a robust hybrid lithium niobate microring laser source for lithium niobate photonic integrated circuits.

Figure 1(a) shows a diagram of our proposed electrically driven compact hybrid lithium niobate microring laser, which is composed of a commercial CoS laser diodes and a high-Q Er: TFLN microring. The emitter size of the CoS laser diodes is 4 μm × 1 μm, and the input size of taper waveguide of the Er: TFLN microring is 4 μm × 0.5 μm. The coupling loss is about 10 dB due to the mismatch of spot size and facet reflectivity. The on-chip Er: TFLN microring is fabricated on a 500-nm-thickness Z-cut LN thin film with an $Er^{3+}$-doping concentration of 1 mol%, more fabrication detail can be found in Ref. 17. Figure 1(b) show the top view of fabricated compact hybrid lithium niobate microring laser, the perimeter of Er: TFLN microring is about 1 mm. As shown in Figure 1(c), the input port of Er: TFLN microring should align with the output port of CoS laser diodes to achieve efficient light coupling by a 6-Axis alignment system with the resolution of 10 nm. To achieve a stable and compact bonding, the ultraviolet (UV) glue is applied by glue dispenser to fasten the two chips by irradiation with UV light.

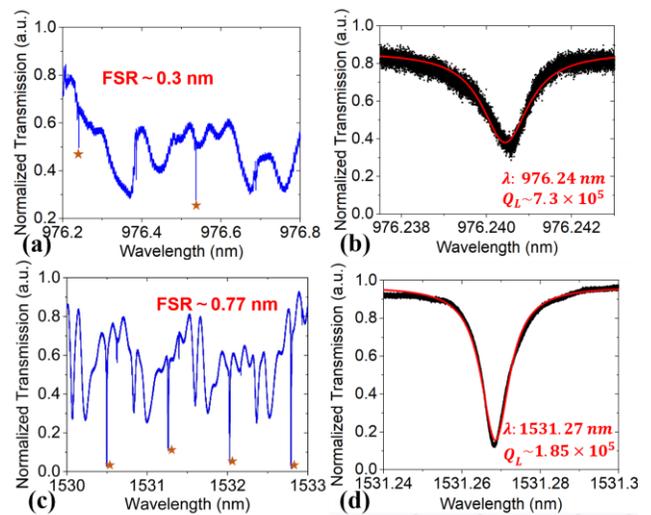

Fig. 2 (a) Transmission spectrum of the Er: TFLN microring laser around the pump laser. (b) The Lorentz fitting (red curve) reveals a Q-factor of $7.3 \times 10^5$ at the wavelength of 976 nm. Transmission spectrum of the mirroring laser around the emission laser. (b) The Lorentz fitting (red curve) reveals a Q-factor of $1.85 \times 10^5$ at the wavelength of 1531 nm.

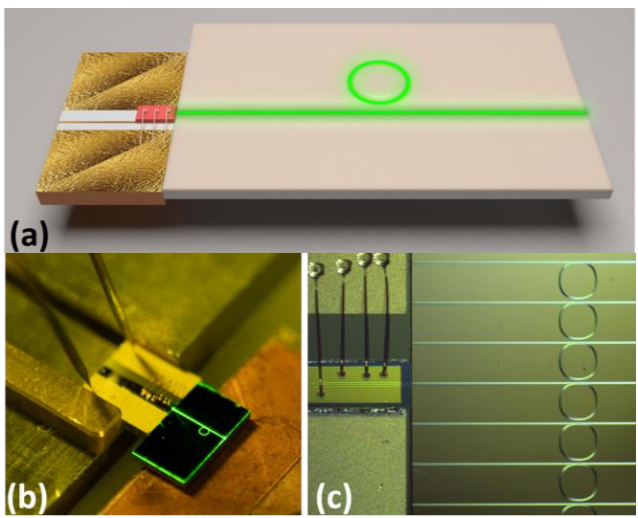

Fig. 1. (a) Scheme of the electrically driven hybrid laser, which is composed of a commercial chip on submount laser diodes and a high-Q Er: TFLN microring laser. (b) Microscope image of of the proposed electrically driven compact hybrid lithium niobate microring laser. (c) Close-up optical micrograph of the interface between the CoS laser diode and Er: TFLN microring.

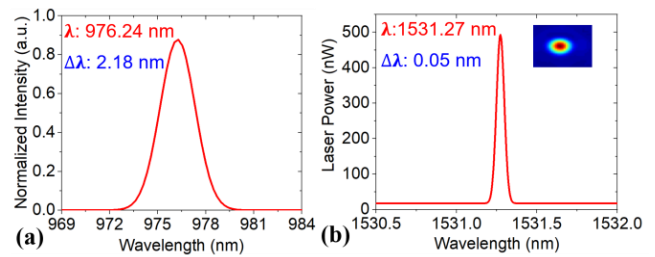

Fig. 3 (a) The spectrum around the pump laser. (b) The spectrum around the emission laser confirming a single mode operation at 1531.27 nm and featuring a linewidth of 0.05 nm, the inset displays infrared imaging of the output of the lasing emission

To characterize the transmission spectrum and Q factors of the Er: TFLN microring, light produced by a continuously tunable laser (CTL 1550, TOPTICA Photonics Inc.) and a power meter (New focus 1811-FC-AC, Newport Inc.) was used for monitoring the power of the light coupled out from the microring laser. Figure 2(a) and (c) show the transmission spectrum of the microring laser, the free spectral range (FSR) of Er: TFLN microring are about 0.3 nm and 0.77nm at the wavelengths around 976 nm and 1531 nm which corresponds to the resonant wavelength of the pump laser and the signal laser, respectively. The Lorenz fitting curves shown in Figure 2(b) and (d) indicate that the microring laser has Q factors $7.3 \times 10^5$ and $1.85 \times 10^5$ at the wavelengths of 976 nm (pump laser) and 1531 nm (lasing emission), respectively.

The lasing emission together with the residual pump laser were coupled out of the microring laser using a lensed fiber. An optical spectrum analyzer (OSA: AQ6375B, YOKOGAWA Inc.) was used to analyze the spectrum of the pump laser and the signal laser, which is shown in Figure 3(a) and (b). The spectrum around the emission laser confirming a single frequency operation at 1531.27 nm and featuring a linewidth of 0.05 nm which is limited by the resolution of the OSA (~0.01 nm). And the linewidth of lasing emission is two orders of magnitude narrower compared with the linewidth of the pump light produced by the CoS laser diodes. The inset of Figure 3(b) shows the single mode output of the lasing emission form the waveguide. Figure 4 (a) shows the pump power dependence of the microring laser. The on-chip laser power as a function of the on-chip pump power is illustrated in Figure 4(b), the lasing threshold was found to be approximately 6 mW by linear fitting, and the conversion efficiency is calculated as $3.9 \times 10^{-3}$ %. Figure 4(c) shows the on-chip laser power as a function of the driving electric power and the threshold current is 0.5 A when operating voltage 1.64 V.

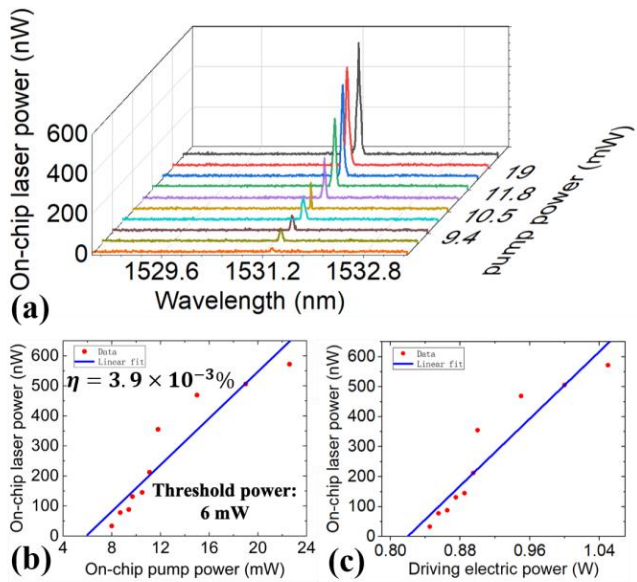

Fig.4. (a) The spectra of the microring laser at the increasing pump powers. (b) The dependence of the on-chip lasing power of the microring laser on the input pump power. (c) The dependence of the on-chip lasing power of the microring laser on the driving electric power.

In conclusion, we demonstrate an electrically driven compact hybrid lithium niobate microring laser by butt coupling a commercial 980-nm pump laser diode chip with a high quality $Er^{3+}$-doped lithium niobate microring chip. Single mode C-band lasing emission form the $Er^{3+}$-doped lithium niobate microring can be observed with the integrated 980-nm laser pumping. The compact hybrid lithium niobate microring laser structure with chip size of 3 mm × 4 mm × 0.5 mm. The threshold pump laser power is 6 mW and the threshold current is 0.5 A (operating voltage 1.64 V) in the atmospheric temperature. The spectrum is single mode and small linewidth of 0.05 nm is observed. This work explores a robust hybrid lithium niobate microring laser source which has potential applications in coherent optical communication and precision metrology.

**Funding.** National Key R&D Program of China (2019YFA0705000), National Natural Science Foundation of China (Grant Nos. 12004116, 12104159, 11874154, 11734009, 11933005, 12134001, 61991444), Science and Technology Commission of Shanghai Municipality (NO.21DZ1101500), Shanghai Municipal Science and Technology Major Project (Grant No.2019SHZDZX01), Shanghai Sailing Program (21YF1410400) and Shanghai Pujiang Program (21PJ1403300).

**Disclosures.** The authors declare no conflicts of interest

**Data Availability.** Data underlying the results presented in this paper are not publicly available at this time but may be obtained from the authors upon reasonable request.